\documentclass[10pt]{amsart}
\usepackage{amsmath,amsthm}
\usepackage[colorlinks=true,linkcolor=blue,urlcolor=blue,citecolor=blue]{hyperref}
\usepackage[all]{xy}

\usepackage{color}

\def\giorno{26/11/2018}

\def\ga{\gamma}

\def\De{\Delta}

\def\s{\sigma}
\def\om{\omega}
\def\vphi{\varphi}
\def\eps{\varepsilon}

\def\pa{\partial}
\def\d{{\mathrm d}}
\def\T{\mathrm{T}}

\def\^#1{\widehat{#1}}

\def\wt#1{\widetilde{#1}}

\def\({\left(}
\def\){\right)}
\def\[{\left[}
\def\]{\right]}

\def\beq{\begin{equation}}
\def\eeq{\end{equation}}

\def\EOR{\hfill $\odot$}

\def\beql#1{\begin{equation} \label{#1}}

\def\eqref#1{(\ref{#1})}

\begin{document}


\title[Symmetry of Stochastic Equations]{Recent advances in symmetry of \\ stochastic differential equations}
\author{Giuseppe Gaeta, Claudia Lunini, Francesco Spadaro}

\date{\giorno}

\begin{abstract}
\noindent We discuss some recent advances concerning the symmetry
of stochastic differential equations, and on particular the
interrelations between these and the integrability -- complete or
partial -- of the equations.
\end{abstract}

\maketitle

{ } \hfill {\it To Gianfausto on his $85^{th}$ birthday}

\section*{Introduction}

The modern theory of Symmetry was laid down by Sophus Lie
(1842-1899). The motivation behind the work of Lie was not in pure
Algebra, but instead in the effort to \emph{solve differential
equations}. This was highly successful; so the question we want to
answer is: can we do something similar for \emph{stochastic}
differential equations?

In this short note we first sketch how the theory of symmetry
helps in determining solutions of (deterministic) differential
equations, both ODEs and PDEs; we will be staying within the
classical theory (Lie-point symmetries), work in coordinates, and
only consider continuous symmetries
\cite{AVL,CGb,KrV,Olv1,Olv2,Ste}. We will then discuss the recent
extension of this theory to \emph{stochastic (ordinary)
differential equations}.

\section{Symmetry of deterministic equations}

\subsection{The Jet space}
\label{sec:jet}

The key idea for a proper treatment of symmetry of (deterministic)
differential equations goes back to E. Cartan and Ch. Ehresmann.
It consists in the introduction of the \emph{jet
bundle} (or jet space if we deal with problems in Euclidean
framework) \cite{ArnGMDE,Olv1,Olv2,Sharpe}.

We denote as \emph{phase bundle} (or \emph{phase space}) the
manifold of dependent ($u^1, ..., u^p$) and independent
($x^1, ..., x^q$) variables; this is naturally seen as a bundle
(with the manifold $B$ where the independent variables live as the
basis) $(M,\pi_0,B)$.

The Jet bundle (of order $n$) $J^n M$ is then the space of
dependent ($u^1,...,u^p$) and independent ($x^1,...,x^q$)
variables, together with the partial derivatives (up to order $n$)
of the $u$ with respect to the $x$; this has also a natural
structure of fiber bundle, $(J^n M , \pi_n , B)$.

We should however keep into account that the $u^a_J$ represents
derivatives of the $u^a$ w.r.t. the $x^i$. In order to do this,
the jet space should be equipped with an additional structure, the
\emph{contact structure} \cite{ArnGMDE,Sharpe}.

This can be expressed by introducing the one-forms $$ \om^a_J \ :=
\ \d u^a_J \ - \ \sum_{i=1}^q \, u^a_{J,i} \, \d x^i \ , $$ which
are called the \emph{contact forms}, and looking at their kernel.

We also mention that moreover, each Jet $J^n M$ is also a fiber
bundle over Jets of lower order; that is, we also have bundles
$(J^n M, \pi_{n,k} , J^k M)$ for all $0 < k < n$. This is at the
basis of the  recursive construction of prolongations of vector
fields (see below).

Jets are a natural generalization of the familiar geometric
description of vector fields: a vector at a given point $x \in M$
can be seen as an equivalence class of curves in $M$ (mutually
tangent at $x$), and a vector field as the choice of a vector at
each point, and as a section of the tangent bundle $T M$. In the
same way, a jet of order $k$ at a given point $x \in M$ can be
seen as an equivalence class of curves in $M$ (mutually tangent of
order $k$ at $x$), and a jet field as the choice of a jet at each
point, and as a section of the jet bundle $J^k M$, with $J^1 M = T
M$, $J^{k+1} M = T (J^k M)$.  We refer e.g. to
\cite{AVL,Olv1,Olv2,Saunders, Sharpe} for  further detail on Jet
bundles and their Geometry.

\subsection{Geometry of differential equations, contact structure,
prolongation}

A differential equation $\De$ determines a manifold in $J^n M$,
the \emph{solution manifold} $S_\De \subset J^n M$ for $\De$. This
is a \emph{geometrical} object; the differential equation can be
identified with it, and we can apply geometrical tools to study
it.

An infinitesimal transformation of the $x$ and $u$ variables is
described by a \emph{vector field} in $M$; once this is defined
the transformations of the derivatives are also implicitly
defined.

The procedure of extending a vector field in $M$ to a vector field
in $J^n M$ by requiring the preservation of the contact structure
-- thus so that derivatives transform in the natural way once the
transformations of dependent and independent variables are given
-- is also called \emph{prolongation}
\cite{AVL,CGb,KrV,Olv1,Olv2,Ste}.

If the vector field $X$ on $M$ is expressed in the local
coordinates $(x,u)$ as $$ X \ = \ \xi^i (x,u) \, \frac{\pa}{\pa
x^i} \ + \ \vphi^a (x,u) \, \frac{\pa}{\pa u^a} \ , $$ its $n$-th
order prolongation $X^{(n)}$ on $J^n M$ is written -- in the local
coordinates $(x,u^{(n)} )$ and in multi-index notation -- as
$$ X^{(n)} \ = \ \xi^i (x,u) \, \frac{\pa}{\pa x^i} \ + \
\psi^a_{J} (x,u^{(|J|)} ) \, \frac{\pa}{\pa u^a_J} \ ; $$ the
coefficients $\psi^a_J$ are provided (recursively) by the
\emph{prolongation formula}
$$ \psi^a_{J,i} \ = \ D_i \psi^a_J \ - \ u^a_{J,k} \ D_i \xi^k
\ ; \ \ \psi^a_0 \ = \ \vphi^a \ . $$

\subsection{Symmetry}

A vector field $X$ defined in $M$ is then a \emph{symmetry} of
$\De$ if its prolongation $X^{(n)}$, satisfies $$ X^{(n)} \ : \
S_\De \ \to \ \T S_\De \ . $$ Note this is a (geometrical)
relations among geometrical objects -- a vector field and a
manifold -- and is hence independent of our choices of
coordinates: as we expect, symmetries will still be present (or
absent) if we change variables.

An equivalent characterization of symmetries is to map solutions
into (generally, different) solutions. In the case a solution is
mapped into itself, we speak of an \emph{invariant solution}.

A first use of symmetry can be that of \emph{generating new
solutions from known ones}. For example, acting with (nontrivial)
symmetries, the solution $u=0$ to the heat equation get
transformed into the fundamental (Gauss) solution; see e.g.
Chapter 3 in \cite{Olv1}.

As we will see, this is by far not the only way in which knowing
(all or some of) the symmetries of a differential equation can
help in determining (all or some of) its solutions.

In order to use the symmetries of a differential equation, we
should of course first of all know what these symmetries are, i.e.
determine them. Determining the symmetry of a given differential
equation goes through the solution of a system of coupled
\emph{linear} PDEs, known indeed as the \emph{determining
equations}.

The procedure for solving them is in general algorithmic and can
be implemented via computer algebra; the exception here is the
case of (systems of) first order ODEs, i.e. Dynamical Systems.

\subsection{Using the symmetry}

The key idea is the same for ODEs and PDEs, and amounts to the use
of \emph{symmetry adapted coordinates}. But the scope of the
application of symmetry methods is rather different in the two
cases, and thus so is the actual meaning of ``adapted''. We will
only consider scalar equations for ease of discussion.

\subsubsection{Symmetry and ODEs}

If an ODE $\De$ of order $n$ admits a Lie-point symmetry $X$, the
equation can be \emph{reduced} to an equation of order $n-1$. The
solutions to the original and to the reduced equations are in
correspondence through a quadrature (which of course introduces an
integration constant).

The main idea is to change variables $(x,u) \to (y,v)$, so that in
the new variables the symmetry vector field $X$ reads
$$ X \ = \ \pa \, / \, \pa v \ . $$
As \emph{$X$ is still a symmetry}, this means that the equation
will not depend on $v$, only on its derivatives.

At this point, with a new  change of coordinates $w  :=  v_y $ we
reduce the equation to one of lower order.

A solution $w = h(y)$ to the reduced equation identifies solutions
$v = g(y)$ to the original equation (in ``intermediate''
coordinates) simply by integrating, $$ v (y) \ = \ \int w(y) \, d
y \ ; $$ a constant of integration will appear here. Finally go
back to the original coordinates inverting the first change of
coordinates.

Note that the reduced equation could still be too hard to solve.
That is, the method can only guarantee that we are reduced to a
problem of lower order, i.e. hopefully simpler than the original
one.

If we are able to solve this reduced problem, then solutions to
the original and the reduced problem are in (many to one)
correspondence.

This approach extends, with certain algebraic conditions, to the
case where multiple symmetries are present, and correspondingly
multiple reductions are possible -- at least if the symmetry
vector field span a solvable Lie algebra \cite{Olv1}.

\subsubsection{Symmetry and PDEs}

The approach in the case of PDEs is in a way at the opposite as
the one for ODEs. If $X$ is a symmetry for $\De$, we change
coordinates $ (x,t;u) \, \to \, (y , s ; v ) $ so that in the new
coordinates $$ X \ = \ \frac{\pa}{\pa y} \ . $$

Now our goal will \emph{not} be to obtain a general reduction of
the equation, but instead to obtain a (reduced) equation which
determines the \emph{invariant solutions} to the original
equation.\footnote{The reason for this is quite clear: changing
coordinates so that the vector field is written as $X = (\pa / \pa
v)$, as in the ODE case, would lead to an equation not explicitly
depending on $v$ (as in the ODE case), i.e. an equation in which
only derivatives of $v$ appear (as in Hamilton-Jacobi). But as
different partial derivatives are present, we cannot reduce the
order of the equation and thus, in general, have no real advantage
by such a transformation.}

In the new coordinates, this is just obtained by \emph{imposing}
$v_y = 0$, i.e. $v = v(s)$. The reduced equation will have (one)
less independent variables than the original one.

This reduced equation will \emph{not} have solutions in
correspondence with general solutions to the original equation:
only the invariant solutions will be common to the two equations.
Contrary to the ODE case, we do not need to solve any
``reconstruction problem''.

\medskip\noindent
{\bf Remark 1.} It was shown by Kumei and Bluman \cite{KBl} that
the (algorithmic) symmetry analysis is also able to detect if a
nonlinear equation can be linearized by a change of coordinates.
The reason is that the underlying linearity will show up through a
Lie algebra reflecting the \emph{superposition principle}.
Similar, albeit more delicate, considerations lead to relating
suitable symmetries and the presence of a \emph{nonlinear
superposition principle} \cite{CGM}.

\medskip\noindent
{\bf Remark 2.} The concept of symmetry was generalized in many
ways; this extends the range of applicability of the theory. We
are not discussing these, but just refer e.g. to
\cite{CGb,KrV,Olv1,Ste}.

\subsubsection{ODEs vs PDEs reduction}

Note that in geometrical terms, the difference between the ODE and
the PDE reduction approach is clearly understood in terms of the
fibration $(M,\pi_0,B)$, see Sect.\ref{sec:jet}: in both cases we
straighten the vector field $X$, but in the ODE case this is done
so that in the new coordinates, and hence in the new fibration
$(M,\wt{\pi}_0,\wt{B})$, $X$ is a vertical vector field; while in
the PDE case this is so that in the new fibration $X$ has no
vertical component.

Correspondingly, in the ODE case $X$ acts on (sections of
$(M,\wt{\pi}_0,\wt{N})$ representing) solutions by parallel
transporting them along fibers; in the new variables $(y,v)$ new
solutions are obtained from known ones, acting with $X$, by the
addition of a constant, which is just the integration constant
arising from the quadrature linking $v(y)$ to $w(y)$.

In the PDE case instead the (sections of $(M,\wt{\pi}_0,\wt{N})$
representing) solutions are invariant when transported
horizontally; this means that the corresponding sections have some
flat directions, and thus depend effectively on a smaller number
of variables than general solutions.

\section{Symmetry of SDEs}
\label{sec:geom}

We will now see how the classical symmetry theory for
(deterministic) differential equations can be extended to the
framework of \emph{stochastic} differential equations.

\subsection{Types of symmetries for SDEs}

We consider an Ito SDE \beql{eq:Ito} d x^i \ = \  f^i (x,t) \, d t
\ + \ \s^i_j (x,t) \, d w^j  \eeq (note by this we always mean a
vector one, i.e. a system of SDEs), and a general vector field
acting in the $(x,t)$ space, \beql{eq:XIto} X \ = \ \tau \, \pa_t
\ + \ \xi^i \, \pa_i \ . \eeq Note that we allow, in general, the
coefficients $\xi^i$ of $X$ to depend on the $(x,t,w)$ variables,
while it makes sense to restrict the dependence of $\tau$ to the
$t$ variable alone \cite{GS17}.

The vector field $X$ in \eqref{eq:XIto} is a symmetry of the Ito
equation \eqref{eq:Ito} if it satisfies the suitable determining
equations; in the general case these are rather involved (see
\cite{GS17} for their explicit expression), and will not be
reported here.

We distinguish different types of symmetries. In particular,
\emph{simple} symmetries act only on the $x$, while \emph{general}
symmetries\footnote{Actually, besides these, also
\emph{W-symmetries} are possible (these also act on the $w^j$),
but will not be considered here. They are characterized by more
general equations, reducing to \eqref{eq:detIto} for vector fields
of the form \eqref{eq:XIto}; see the discussion in \cite{GS17}.}
act on both the $x$ and $t$. We will also distinguish between
\emph{deterministic} symmetries, i.e. those for which -- with
reference to \eqref{eq:XIto} -- we have $\xi = \xi (x,t)$ and
$\tau = \tau (t)$; and \emph{random} symmetries, i.e. those with
$\xi = \xi (x,t,w)$, $\tau = \tau (t)$.

We are specially interested, for reason which will be clear in the
following, in \emph{simple} (possibly random) symmetries, hence in
the case $\tau = 0$ in \eqref{eq:XIto}. In this case the
determining equations for (simple) symmetries of \eqref{eq:Ito}
read \beql{eq:detIto} \begin{cases} \pa_t \xi^i \ + \ f^j \, (\pa_j
\xi^i) \ - \ \xi^j \, (\pa_j f^i ) \ = \ - \frac12 (\triangle
\xi^i ) & , \\ \^\pa_k \xi^i \ + \ \s^j_{\ k} \, (\pa_j \xi^i) \
- \ \xi^j \, (\pa_j \s^i_{\ k} ) \ = \ 0 & . \end{cases} \eeq Here $\^\pa_i
:= \pa / \pa w^i$, and the symbol $\triangle$ denotes the Ito
Laplacian \beq \triangle u \ := \ \sum_{j,k=1}^n \left[ \left( \s
\, \s^T \right)^{jk} \, \frac{\pa^2 u}{\pa x^j \pa x^k} \ + \ 2 \
\s^{ik} \, \frac{\pa^2 u}{\pa x^j \pa w^k} \ + \ \delta^{jk} \,
\frac{\pa^2 u}{\pa w^j \pa w^k} \right]  \ . \eeq

\medskip\noindent
{\bf Remark 3.} The case with $\xi^i = \xi^i (x,t)$ and $\tau
=\tau (x,t)$ would also deserve the name of ``deterministic'', but
it is not acceptable in view of other considerations (roughly
speaking because we want to keep $t$ as a deterministic smooth
variable, while $x$ is in this context a random one, and hence we
should not mix it with $t$); see the discussion in \cite{GS17}.
Similar considerations apply also to the case with $\xi^i
(x,t;w)$, where one would be tempted to consider $\tau = \tau
(x,t;w)$ rather than just $\tau = \tau (t)$.

\subsection{Symmetry of SDEs and change of variables}

When we look at symmetry of a SDEs {\it per se} a substantial
problem is present.

In fact, the symmetry approach is based on passing to
symmetry-adapted coordinates; vector fields transform
``geometrically'' (i.e. via the chain rule) under changes of
coordinates, and \emph{deterministic} differential equations are
(identified with) geometrical objects, hence also transform
geometrically. It is then obvious that \emph{symmetry are
preserved under changes of coordinates}, as already stressed
above.

On the other hand, an Ito equation is \emph{not} a geometrical
object: in fact, it transforms under the Ito rule, not the chain
rule. Thus it is \emph{not} granted that $X$ will still be a
symmetry when we change coordinates so that $X = \pa_x$! Note this
is also true for \emph{deterministic} symmetries of stochastic
equations.

The easy way out of this problem would be giving up Ito equations
and using \emph{Stratonovich equations} instead. These do
transform according to the chain rule, i.e. geometrically; but the
relation between an Ito and the corresponding Stratonovich process
is not that obvious -- especially in this respect \cite{Stroock}.

In fact, it is known that in general the two do \emph{not} share
the same symmetries \cite{Unal}. But it is also known that they
have the same \emph{simple} symmetries, and this both in the
deterministic \cite{Unal} and in the random \cite{GL1} case. This
fact is specially interesting, as the Kozlov theory
\cite{Koz1,Koz2,Koz3} relating symmetry to integrability of SDEs
only makes use of simple symmetries.

We note that the determining equations for simple symmetries --
which we still write in the general form \eqref{eq:XIto} -- of a
Stratonovich equation \beql{eq:Stra} d x^i \ = \ b^i (x,t) \, d t
\ + \ \s^i_{\ k} (x,t) \circ d w^k \eeq turn out to be \cite{GS17}
\beql{eq:symmStra} \begin{cases} \pa_t \xi^i \ + \ b^j (\pa_j \xi^i)
\ - \ \xi^j (\pa_j b^i) \ = \ 0 & , \\ \^\pa_k \xi^i \ + \ \s^j_{\ k}
(\pa_j \xi^i) \ - \ \xi^j (\pa_j \s^i_{\ k} ) \ = \ 0 & . \end{cases}
\eeq (The reader is again referred to \cite{GS17} for the general
case.)

In particular, if we consider the Stratonovich equation associated
to the Ito equation \eqref{eq:Ito}, i.e. for \beq f^i \ = \ b^i \
+ \ \frac12 \, \left[ \frac{\pa (\s^T)^i_{\ j}}{\pa x^k} \right] \
\s^{kj} \ := \ b^i \ + \ \rho^i  \ , \eeq then these determining
equations for simple symmetries read \beql{eq:detStra}
\begin{cases} \pa_t \xi^i \ + \ f^j \, (\pa_j \xi^i) \ - \ \xi^j \,
(\pa_j f^i ) \ = \ \rho^j (\pa_j \xi^i) \ - \ \xi^j (\pa_j \rho^i)
& , \\ \^\pa_k \xi^i \ + \ \s^j_{\ k} \, (\pa_j \xi^i) \ - \
\xi^j \, (\pa_j \s^i_{\ k} ) \ = \ 0 & . \end{cases} \eeq
They appear to be
in general \emph{different} form the determining equations
\eqref{eq:detIto} for the Ito equation.

\subsection{Unal type theorems}

It turns out that, as can be checked by a careful explicit
computation, the difference between \eqref{eq:detIto} and
\eqref{eq:detStra} is only apparent. In fact, we have the
following result, shown by Unal \cite{Unal} for the deterministic
case and then extended to the random one \cite{GL1} (we refer to
the original papers for its proof).

\medskip\noindent
{\bf Proposition 1.} {\it The simple \emph{deterministic or
random} symmetries of an Ito equation and those of the equivalent
Stratonovich equation do coincide.}
\bigskip

In his paper, however, Unal also showed that -- even in the
deterministic framework -- the result does \emph{not} extend to
more general symmetries; in particular, if one considers
symmetries with generator of the general form \eqref{eq:XIto},
thus in general with $\tau \not= 0$, then the determining
equations for the Ito and the associated Stratonovich equation are
equivalent if and only if $\tau$ satisfies the additional
condition \beq \s^k_{\ p} \, \s^{ip} \ \left[ \pa_k \left( \pa_t
\tau \ + \ f^j \, (\pa_j \tau ) \ + \ \frac12 \, \s^m_{\ q} \,
\s^j_{\ q} \,(\pa_m \pa_j \tau) \right) \right] \ = \ 0 \ . \eeq

We stress that this condition is identically satisfied for $\tau =
\tau (t)$, i.e. for ``acceptable'' cases according to the
discussion in \cite{GS17}.

Thus we conclude that for (deterministic or random) simple
symmetries, and actually also for the corresponding ``acceptable''
general symmetries, i.e. with $\tau = \tau (t)$, symmetries of an
Ito equation and of the associated Stratonovich one do coincide.
As the latter are preserved under changes of variables, it follows
that the former are preserved as well. In the end \cite{GL1,GL2},

\medskip\noindent
{\bf Lemma 1.} {\it Simple (random or deterministic) symmetries of
an Ito equation are preserved under changes of coordinates.}

\medskip\noindent
{\bf Remark 4.} This entails that we can hope to use (simple or
acceptable general) symmetries of Stochastic Differential
Equations, as the basic ingredient for applications of the theory
-- i.e. indeed preservation of symmetries under changes of
variables -- is there, albeit in a much less immediate way than
for deterministic differential equations.

\section{Kozlov theory}
\label{sec:Koz}

In the deterministic case, symmetry guarantees that an ODE can be
reduced (or solved). The same holds in the SDE case, but only
\emph{simple} symmetries $X = f^i (x,t) \pa_i$
matter.\footnote{This limitation may look surprising at first, but
one should note that now $x$ and $t$ are intrinsically different:
one is a random process (indexed by $t$), the other a smooth
deterministic variable.}

We have the following theorem, which is due to R. Kozlov
\cite{Koz1} (see also \cite{GL2,Lunini}):

\medskip\noindent
{\bf Theorem 1.} {\it The scalar SDE \beql{eq:K1} d y \ = \ \wt{f}
(y,t) \ d t \ + \ \wt{\s} (y,t) \ d w \eeq can be transformed by a
deterministic map $y = y (x,t)$ into \beql{eq:K2} d x \ = \ f(t)
\, d t \ + \ \s (t) \, d w \ , \eeq and hence explicitly
integrated, \emph{if and only if} it admits a simple deterministic
symmetry.

If the generator of the latter is $ X  = \vphi (y,t) \pa_y$, then
the change of variables $y =  F (x,t)$ transforming \eqref{eq:K1}
into \eqref{eq:K2} is the inverse to the map $x = \Phi (y,t)$
identified by
$$ \Phi (y,t) \ = \ \int \frac{1}{\vphi (y,t) } \ d
y \ . $$}

\medskip\noindent
{\bf Example 1.} The Ito equation $ d y \ = \ \[ e^{-y} - (1/2)
e^{- 2y} \] \, dt \ + \ e^{-y} \, d w $ admits the simple
deterministic symmetry generated by $X = e^{-y} \pa_y$. The
associated change of variable is $ x = e^y$; in terms of this $X =
\pa_x$, and the equation reads $ d x \ = \ dt \ + \ dw \ ,$ which
is readily integrated. \EOR
\bigskip

The same approach can be pursued to study \emph{partial
integrability}, i.e. reduction of an $n$-dimensional SDE to an SDE
in dimension $n-r$ plus $r$ (stochastic) integrations.

In the deterministic case, this is possible \emph{if and only if}
there are $r$ simple symmetry generators spanning a solvable Lie
algebra. In the stochastic case we obtain essentially the same
result, but now it is convenient to consider separately the case
of deterministic symmetries and that of random ones. Again the
relevant results in this direction have been obtained by Kozlov
\cite{Koz2,Koz3} (see also \cite{GL2,Lunini}).

This has been considered in the literature only for (multiple)
deterministic simple symmetries; the result below is quoted
verbatim from \cite{GL2}, and the reader is referred there for the
proof.

\medskip\noindent
{\bf Theorem 2.} {\it Suppose the system \eqref{eq:Ito} admits an
$r$-parameter solvable algebra $\mathcal{G}$ of simple
deterministic symmetries, with generators \beq
\mathbf{X}_k=\sum_{i=1}^n \varphi_k^i(x,t) \ \frac{\pa}{\pa x_i}
\qquad (k=1,...,r) \ , \eeq acting regularly with $r$-dimensional
orbits.

Then it can be reduced to a system of $m = (n-r)$ equations, \beq
d y^i \ = \ g^i (y^1,...,y^m;t) \, d t \ + \ \s^i_{\ k}
(y^1,...,y^m;t) \, d w^k \ \ \ \ (i,k=1,...,m) \eeq and $r$
``reconstruction equations'', the solutions of which can be
obtained by quadratures from the solution of the reduced
$(n-r)$-order system. In particular, if $r=n$, the general
solution of the system can be found by quadratures.}
\bigskip

We note that in Kozlov's original paper \cite{Koz2} (see Example
4.2 in there) the Theorem is applied to any linear two-dimensional
system of SDEs \beq \begin{cases} d x_1 \ = \ (a_1 \, + \, b_{11} \, x_1
\, + \, b_{12} \, x_2 ) \, d t \ + \ s_{11} \,  d w_1 \ + \ s_{12}
\, d w_2 & , \\
 d x_2 \ = \ ( a_2 \, + \, b_{21} \, x_1 \, + \, b_{22} \, x_2 ) \, dt \ + \
s_{21} \, d w_1 \ + \ s_{22} \ , d w_2 & ; \end{cases} \eeq see there for a detailed discussion and results.

We have so far only considered deterministic symmetries. In the
case of random symmetries, the associated random change of
variables could change the Ito equation into a random system of
different nature. This problem accounts for the appearance of an
extra condition, absent when one is only considering deterministic
simple symmetries.

\medskip\noindent
{\bf Theorem 3.} {\it Let the Ito equation \beql{eq:dyRR} d y \ =
\ F(y,t) \, d t \ + \ S (y,t) \, d w \eeq admit as Lie-point
symmetry the simple random vector field \beql{eq:Xrvf} X \ = \
\vphi (y,t,w) \, \pa_y \ . \eeq If there is a determination of
\beql{eq:vphiPhi} \Phi (y,t,w) \ = \ \int \frac{1}{\vphi (y,t,w)}
\ d y \eeq such that the equations \beql{eq:dsfw0} \Phi_{ww} \ + \
S \, \Phi_{yw} \ = \ 0 \ ; \ \ \ \ \ \Phi_{tw} \ + \ F \,
\Phi_{yw} \ + \ (1/2) \, \( \Delta \Phi \)_w \ = \ 0 \eeq are
satisfied, then the equation is reduced to the explicitly
integrable form \beql{eq:dxRint} d x \ = \ f(t) \, d t \ + \ \s
(t) \, d w \eeq by passing to the variable $x = \Phi (y,t,w)$.}

\medskip\noindent
{\bf Remark 5.} This formulation is not fully satisfactory, in
that it is based on the existence of a determination of an
integral with certain properties. One would like to have a
criterion based on the directly available data, i.e. the functions
$F(y,t)$, $S(y,t)$ and $\vphi (y,t,w)$. This is provided by the
next Theorem.

\medskip\noindent
{\bf Theorem 4.} {\it Let the Ito equation \eqref{eq:dyRR} admit
as Lie-point symmetry the simple random vector field
\eqref{eq:Xrvf}; define $\ga (y,t,w) := \pa_w ( 1 / \vphi )$.

If the functions $F(y,t)$, $S(y,t)$ and $\ga (y,t,w)$ satisfy the
relation \beql{eq:bcomp} S \, \ga_{t} \ + \ S_t \, \ga \ = \ F \,
\ga_{w} \ + \ (1/2) \, \[ S \, \ga_{ww} \ + \ S^2 \, \ga_{yw} \] \
, \eeq then the equation \eqref{eq:dyRR} can be mapped into an
integrable Ito equation \eqref{eq:dxRint} by a simple random
change of variables.}

\medskip\noindent
{\bf Example 2.} The equation $dy \ = \ y \, e^{-t} \, dt \ + \ y
\, dw $ admits the simple random symmetries $X = \eta (\zeta)
\pa_y$, with $\eta$ an arbitrary function of $\zeta = 2 e^{-t} + t
- w + \log (y)$; equation \eqref{eq:bcomp} is satisfied. Let us
choose for definiteness $\eta (\xi) = \xi$. The associated change
of variable is then $ x = (1/2) \log [2 + e^t (2 - w) + 2 e^t \log
(y)] + \beta (t,w)$; the resulting equation is of Ito form for
$\beta (t,w) = b(t) + c w$, with $b$ an arbitrary function. Then
we get $ d x \ = \ [ b' (t) \ + \ (1/2) ] \, dt \ + \ c \, dw $.
\EOR

\medskip\noindent
{\bf Example 3.} The equation $ dy \ = \ dt \ + \ y \, dw $ has
the simple random symmetry $X = \exp [w-t/2] \pa_y$; the
associated new variable is $x = \exp [t/2 - w] + \beta (t,w)$ and
in this case eq.\eqref{eq:bcomp} is \emph{not} satisfied, for any
choice of $\beta$. In term of this the equation reads $ dx \ = \
\exp [ t/2 - w] \, dt \ ; $ this is \emph{not} in Ito form, but is
readily integrated. \EOR


\medskip\noindent
{\bf Remark 6.} The Theorems 3 and 4 identify the presence of a
simple random symmetry $X = \vphi^i (x,t;w) \pa_i$ such that the
compatibility condition \eqref{eq:bcomp} is satisfied as a
\emph{sufficient} condition for integrability. It is quite simple
to observe this is also a \emph{necessary} condition.

\medskip\noindent
{\bf Theorem 5.} {\it Let the Ito equation \eqref{eq:dyRR} be
reducible to the integrable form \eqref{eq:dxRint} by a simple
random change of variables $x = \Phi (y,t;w)$. Then necessarily
\eqref{eq:dyRR} admits \beq X \ = \ \[ \Phi_y (y,t,w) \]^{-1} \,
\pa_y \ := \ \varphi (y,t,w) \, \pa_y \eeq as a symmetry vector
field, and -- with $\ga = \pa_w (1/\vphi)$ -- \eqref{eq:bcomp} is
satisfied.}
\bigskip

\section{Discussion and conclusions}

The symmetry approach is a general way to tackle Differential
Equations; in the deterministic framework it proved invaluable
both for the theoretical study of differential equations and for
obtaining their concrete solutions. The theory is comparatively
much less advanced in the case of stochastic differential
equations. A first obstacle lies in that it is not at all obvious
that symmetries of SDEs are preserved under changes of variables;
this is the case for a special class of symmetries -- 
the one of interest for concrete applications -- as discussed
in Section \ref{sec:geom}.

There is now some general agreement on what the ``right'' (that
is, useful) definition of symmetry for SDE is; but only few
applications have been considered, most of these concerning
integrable or partially integrable equations.

Theorems equivalent to the standard ones for ODEs have been
obtained by R. Kozlov -- and recently extended -- for (ordinary)
SDEs, both for what concerns \emph{solving} equations and for
\emph{reducing them}; these have been discussed in Section
\ref{sec:Koz}. The big difference with respect to the
deterministic case is that now we cannot use \emph{general}
symmetries, but only \emph{simple} ones.

Even beside this, there is undoubtedly ample space for considering
new applications, first and foremost considering ``non
integrable'' equations. Correspondingly, there is ample space for
concrete applications, i.e. applying the approaches already
existing or to be developed to new concrete stochastic systems.
\medskip

We conclude by a number of observations:

\medskip\noindent
$(i)$ An important topic has been completely absent from our
discussion: that is, \emph{symmetry of variational problems}
(Noether theory). For this we refer e.g. to \cite{Mis,Yas,Zam}.

\medskip\noindent
$(ii)$ Similarly, we have not discussed the interrelations between
symmetries of an Ito equation and those of the associated
diffusion (Fokker-Planck) equation; for this we refer e.g. to
\cite{GPR,GRQ}.

\medskip\noindent
$(iii)$ Reduction by multiple symmetries has been studied in the
literature only in the case of deterministic symmetries. Albeit it
appears that no obstacle is present in the case of random
symmetries -- except that, as for a single symmetry, the
compatibility condition studied above should also be required --
one would like to have precise statements in this respect.

\medskip\noindent
$(iv)$ In the deterministic framework, symmetry theory flourished
and expanded its role by considering generalization of the
``standard'' (i.e. Lie-point) symmetries in several directions
\cite{AVL,KrV,Olv1,Ste}. As far as we know, there is no attempt in
this direction for stochastic systems yet; any work in this
direction is very likely to collect success and relevant results.

\medskip\noindent
$(v)$ Also, so far only \emph{first order systems} have been
considered; but Physical applications often require to consider
\emph{second order} ones (as in the familiar case of
Einstein-Smoluchowsky {\it vs.} Ornstein-Uhlenbeck processes).
This is definitely a direction requiring serious investigation,
also in connection with the previous points.\footnote{In
particular, considering second (or higher) order equations opens
the way to introduction of twisted prolongations and twisted
symmetries \cite{Gtwist}.}



\section*{Appendix. Derivation of the determining equations}

In this Appendix we briefly discuss how the determining equations
\eqref{eq:detIto} and \eqref{eq:detStra} are obtained; see e.g.
\cite{GPR,GL1,GL2,GS17} for further detail.

The vector field $X = \xi^i (x,t) \pa_i$ generates an
infinitesimal map $x^i \to x^i + \eps \xi^i (x,t)$; under this the
different objects appearing in \eqref{eq:Ito} map as follows (all
functions depend on $x,t$):
\begin{eqnarray*}
f^i &\to& f^i  \ + \ \eps \, [\pa f^i  / \pa x^j] \, \xi^j  \ , \
\ \ \
\s^i_{\ k} \ \to \ \s^i_{\ k} \ + \ \eps \, [ \pa \s^i_{\ k} / \pa x^j ] \, \xi^j \ ; \\
d x^i &\to& d x^i \ + \ \eps \, d \xi^i \\ & & \ = \ d x^i \ + \ \eps \, \[ (\pa \xi^i / \pa t) \, d t \ + \
(\pa \xi^i / \pa x^j) \, d x^j \ + \ (1/2) (\Delta \xi^i ) \, d t \] \ ; \end{eqnarray*}
note that the last term in the last line originates from Ito formula. Plugging these into \eqref{eq:Ito}, and requiring that this is actually mapped into itself -- i.e. the vanishing of terms of order $\eps$ -- we obtain exactly the determining equations \eqref{eq:detIto}.

In the case of Stratonovich equations \eqref{eq:Stra} we proceed
in the same way, but now variables do \emph{not} change according
to the Ito rule, following instead the usual chain rule. Thus in
this case
\begin{eqnarray*}
b^i &\to& b^i  \ + \ \eps \, [\pa b^i  / \pa x^j] \, \xi^j  \ , \
\ \ \ 
\s^i_{\ k} \ \to \ \s^i_{\ k} \ + \ \eps \, [ \pa \s^i_{\ k} / \pa x^j ] \, \xi^j \ ; \\
d x^i &\to& d x^i \ + \ \eps \, d \xi^i \ = \ d x^i \ + \ \eps \,
\[ (\pa \xi^i / \pa t) \, d t \ + \ (\pa \xi^i / \pa x^j) \, d x^j
\] \ , \end{eqnarray*} with no Ito term in the $d x^i$ change. We
plug these into the Stratonovich equation \eqref{eq:Stra} and
require it is mapped into itself -- i.e. that terms of order
$\eps$ vanish -- and thus obtain the determining equations
\eqref{eq:symmStra}.

\section*{Acknowledgements}

A relevant part of this work was performed in the stay of GG at
SMRI; FS is supported by the CONSTAMIS ERC grant. We thank the
Referees for constructive criticism, and the Editors for allowing
extra space to meet these requirements.



\vfill
{\small
\par\noindent
G. Gaeta: {\it Dipartimento di Matematica, Universit\`a degli Studi di Milano, v. Saldini 50, I-20133 Milano (Italy);
\tt{giuseppe.gaeta@unimi.it}};
\par\noindent
C. Lunini: {\it Istituto di
Tecnologie Industriali e Automazione - CNR, Via G. Previati 1E,
I-23900 Lecco (Italy); {\tt cla.lunini@icloud.com}};
\par\noindent
F. Spadaro: {\it EPFL-SB-MATHAA-CSFT, Batiment MA - Station 8,
CH-1015 Lausanne (Switzerland); {\tt francesco.spadaro@epfl.ch}}
}


\begin{thebibliography}{39}


\bibitem{AVL} Alekseevsky D.V., Vinogradov A.M. \& Lychagin V.V.,
{\it Basic ideas and concepts of Differential Geometry}, Springer
1991

\bibitem{ArnGMDE} Arnold V.I., {\it Geometrical Methods in the
Theory of Ordinary Differential Equations}, Springer 1983

\bibitem{CGM} Carinena J.F., Grabowski J. \& Marmo G., {\it
Lie-Scheffers systems. A geometric approach}, Bibliopolis 2000

\bibitem{CGb} Cicogna G. \& Gaeta G., {\it Symmetry and perturbation theory in
nonlinear dynamics}, Springer 1999

\bibitem{Gtwist} Gaeta G., ``Twisted symmetries of differential
equations'', {\it J. Nonlin. Math. Phys.} {\bf 16} (2009),
S107-S136; ``Simple and collective twisted symmetries'', {\it J.
Nonlin. Math. Phys.} {\bf 21} (2014), 593-627

\bibitem{GPR} Gaeta G., ``Symmetry of stochastic non-variational
differential equations'', {\it Phys. Rep.} {\bf 686} (2017), 1-62
[Erratum: {\bf 713}, 18]


\bibitem{GL1} Gaeta G. \& Lunini C., ``On Lie-point symmetries for Ito
stochastic differential equations'', {\it J. Nonlin. Math. Phys.}
{\bf 24-S1} (2017), 90-102

\bibitem{GL2} Gaeta G. \& Lunini C., ``Symmetry and integrability for
stochastic differential equations'', {\it J. Nonlin. Math. Phys.}
{\bf 25} 2018, 262-289

\bibitem{GRQ} Gaeta G. \& Rodr\'iguez-Quintero N., ``Lie-point
symmetries and stochastic differential equations'', {\it J. Phys.
A} {\bf 32} (1999), 8485-8505; ``Lie-point symmetries and stochastic differential equations: II'' {\it J. Phys. A} {\bf 33} (2000), 4883-4902

\bibitem{GS17} Gaeta G. \& Spadaro F., ``Random Lie-point symmetries
of stochastic differential equations'', {\it J. Math. Phys.} {\bf
58} (2017), 053503 [Erratum, {\it J. Math. Phys. } {\bf 58}
(2017), 129901]

\bibitem{Koz1} Kozlov R.,
``The group classification of a scalar stochastic differential
equation'', {\it J. Phys. A} {\bf 43} (2010), 055202

\bibitem{Koz2} Kozlov R., ``Symmetry of systems of stochastic
differential equations with diffusion matrices of full rank'',
{\it J. Phys. A} {\bf 43} (2010), 245201

\bibitem{Koz3} Kozlov R.,
``On maximal Lie point symmetry groups admitted by scalar
stochastic differential equations'', {\it J. Phys. A} {\bf 44}
(2011), 205202

%

\bibitem{KrV} Krasil'schik I.S. \&  Vinogradov A.M.,
{\it Symmetries and conservation laws for differential equations
of mathematical physics}, A.M.S. 1999

\bibitem{KBl} Kumei S. \& Bluman G., ``When nonlinear differential
equations are equivalent to linear differential equations'', {\it
SIAM J. Appl. Math.} {\bf 42} (1982), 1157-1173

\bibitem{Lunini} Lunini C., ``Symmetry approach to the integration of
stochastic differential equations'', {\it M.Sc. Thesis},
Universit\`a degli Studi di Milano, 2017

\bibitem{Mis} Misawa T., ``Noether's theorem in symmetric stochastic
calculus of variations'', {\it J. Math. Phys.} {\bf 29} (1988),
2178-2180

\bibitem{Olv1} Olver P.J., {\it Application of Lie groups to
differential equations}, Springer 1986

\bibitem{Olv2} Olver P.J., {\it Equivalence, Invariants,
and Symmetry}, Cambridge UP 1995

\bibitem{Saunders} Saunders D.J. , {\it The Geometry of Jet Bundles},
Cambridge UP 1989

\bibitem{Sharpe} Sharpe R.W., {\it Differential Geometry},
Springer 1997

\bibitem{Ste} Stephani H., {\it Differential equations. Their
solution using symmetries}, Cambridge UP 1989

\bibitem{Stroock} Stroock D.W.,  {\it Markov processes from K.Ito's
perspective}, Princeton UP 2003

\bibitem{Unal}  Unal G., ``Symmetries of Ito and Stratonovich Dynamical
Systems and Their Conserved Quantities'', {\it Nonlinear Dynamics}
{\bf 32} (2003), 417-426

\bibitem{Yas} Yasue K., ``Stochastic calculus of variations'', {\it Lett.
Math. Phys.} {\bf 4} (1980), 357-360; ``Stochastic calculus of variations'', {\it J. Funct. Anal.} {\bf 41} (1981), 327-340

\bibitem{Zam} Zambrini J.C., ``Stochastic dynamics: A review of
stochastic calculus of variations'', {\it Int. J. Theor. Phys.}
{\bf 24} (1985), 277-327

\end{thebibliography}
\end{document}